\documentclass[%
 reprint,
superscriptaddress,
nofootinbib,
 amsmath,amssymb,
 aps,
 prr,
longbibliograohy,
]{revtex4-2}
\usepackage{bbold}
\usepackage[breaklinks=true,colorlinks=true
,urlcolor=blue
,anchorcolor=blue
,citecolor=blue
,filecolor=blue
,linkcolor=blue
,menucolor=blue
,pagecolor=blue
,linktocpage=true
,pdfproducer=medialab
,pdfa=true
]{hyperref}
\usepackage{graphicx,hyperref,xcolor}

\definecolor{rossocorsa}{rgb}{0.83, 0.0, 0.0}
\definecolor{bleudefrance}{rgb}{0.19, 0.55, 0.91}
\usepackage{hyperref}
\hypersetup{
    colorlinks,
    citecolor=bleudefrance,
    filecolor=bleudefrance,
    linkcolor=rossocorsa,
    urlcolor=bleudefrance
}

\usepackage[T1]{fontenc}   
\usepackage[utf8]{inputenc}

\usepackage{color}
\usepackage[dvipsnames,svgnames]{xcolor}
\usepackage{graphicx}
\usepackage{dcolumn}
\usepackage{bm}
\usepackage{bbm}
\usepackage{mathtools}
\usepackage{amsmath,braket}
\usepackage{amssymb,physics}
\usepackage{float}

\usepackage{tikz}
\usepackage{circuitikz}
\usetikzlibrary{shapes.geometric} 
\usetikzlibrary{3d,calc}
\usepackage{pgfplots}
\pgfplotsset{compat=1.17}

\usetikzlibrary{decorations.pathreplacing} 
\usetikzlibrary{decorations.markings}
\tikzset{snake it/.style={decorate, decoration=snake}}

\def\m{{\mu}}

\def\ep{{\epsilon}}

\def\frac#1#2{{#1\over #2}}

\def\s{\sqrt}
\def\i{{\rm i}}

\def\be{\begin{equation}}
\def\ee{\end{equation}}
\def\ba{\begin{eqnarray}}
\def\ea{\end{eqnarray}}
\def\de{\partial}

\def\ti{\tilde}

\def\no{\nonumber \\}

\def\la{\langle}
\def\lb{\rangle}
\def\ep{\epsilon}

\def\vp{\varphi}

\newcommand{\TFD}{|\text{TFD}\rangle}
\newcommand{\DFT}{\langle\overline{\text{TFD}}|}

\allowdisplaybreaks[4]

\usepackage{amsthm}
\newtheoremstyle{break}
  {\topsep}  
  {\topsep}   
  {}          
  {}          
  {\bfseries} 
  {.}         
  {\newline}  
  {} 
\theoremstyle{break}

\usepackage{empheq}

\newcommand{\Ls}{L}
\newcommand{\diff}{\text{d}}
\newcommand{\dS}{\text{dS}}
\newcommand{\GN}{G_\text{N}}
\newcommand{\GNd}{G_\text{N}^{(d+1)}}

\newcommand{\iu}{\text i}
\newcommand{\ex}[1]{\text{e}^{#1}}

\begin{document}
\begin{flushright}
YITP-26-102
\\
\end{flushright}
\title{{\large de Sitter holography from a Lorentzian torus}}

\author{Kosei Fujiki}
\affiliation{Center for Gravitational Physics and Quantum Information, Yukawa Institute for Theoretical Physics, Kyoto University, Kitashirakawa Oiwakecho, Sakyo-ku, Kyoto 606-8502, Japan}

\author{Michitaka Kohara}
\affiliation{Center for Gravitational Physics and Quantum Information, Yukawa Institute for Theoretical Physics, Kyoto University, Kitashirakawa Oiwakecho, Sakyo-ku, Kyoto 606-8502, Japan}

\author{Javier Moreno}
\affiliation{Center for Gravitational Physics and Quantum Information, Yukawa Institute for Theoretical Physics, Kyoto University, Kitashirakawa Oiwakecho, Sakyo-ku, Kyoto 606-8502, Japan}
\affiliation{Departament de Física Quàntica i Astrofísica, Institut de Ciències del Cosmos\\ Universitat de Barcelona, Martí i Franquès 1, E-08028 Barcelona, Spain}

\author{Kotaro Shinmyo}
\affiliation{Center for Gravitational Physics and Quantum Information, Yukawa Institute for Theoretical Physics, Kyoto University, Kitashirakawa Oiwakecho, Sakyo-ku, Kyoto 606-8502, Japan}

\author{Tadashi Takayanagi}
\affiliation{Center for Gravitational Physics and Quantum Information, Yukawa Institute for Theoretical Physics, Kyoto University, Kitashirakawa Oiwakecho, Sakyo-ku, Kyoto 606-8502, Japan}
\affiliation{Inamori Research Institute for Science, 620 Suiginya-cho, Shimogyo-ku, Kyoto 600-8411, Japan}


\begin{abstract}
We show that the quantum entanglement between the conformal field theories (CFTs) living on the future and past boundaries in the de Sitter/conformal field theory (dS/CFT) correspondence can be described by Wick rotating to a geometry with two timelike directions. We propose a new realization of the dS/CFT correspondence in which quantum gravity on this two-time geometry is holographically dual to a CFT defined on a Lorentzian torus. We show that this duality reproduces key features of dS holography, including the dS entropy, correlation functions, and pseudoentropy. Finally, by extending the framework of path-integral optimization, we explain how dS spacetime emerges from the CFT on the Lorentzian torus.

\end{abstract}

\maketitle





\noindent\textbf{1. Introduction}. 

The holographic principle~\cite{tHooft:1993dmi,Susskind:1994vu}---which posits that gravity in a bulk space is dual to a certain quantum system on its boundary---is one of the most promising approaches to understanding the dynamics of quantum gravity. 
It has been remarkably successful for anti-de Sitter (AdS) spaces via the AdS/CFT correspondence~\cite{Maldacena:1997re,Gubser:1998bc,Witten:1998qj}.
However, our understanding of holographic dualities for many other spacetimes remains limited.

A prime example of such open problems is holography for de Sitter (dS) space. Its maximal symmetry makes it fundamental in its own right, while its role in inflationary cosmology ties it directly to the origin of our own universe. Yet the nature of its holographic dual remains unresolved. A direct application of the holographic principle to dS space yields the dS/CFT correspondence~\cite{Strominger:2001pn}. This argues that a $d$ dimensional Euclidean conformal field theory (CFT) is dual to gravity on a $d+1$ dimensional dS space. As opposed to the AdS/CFT, the dual CFTs are known to be non-unitary~\cite{Maldacena:2002vr,Anninos:2011ui,Hikida:2021ese}. 
In the dS/CFT, the dual CFT lives on the spacelike boundary of the dS space. There are also other formulations of dS holography by assuming that the boundary theories live on timelike surfaces. For example, this includes the $T\bar T$ deformation approach~\cite{Gorbenko:2018oov,Silverstein:2024xnr,Chang:2025ays}, the surface state duality~\cite{Miyaji:2015yva}, 
the half dS holography~\cite{Kawamoto:2023nki} and the static patch holography in terms of Double Scaled SYK model~\cite{Susskind:2021esx,Narovlansky:2023lfz}.

Among these approaches, the dS/CFT has advantages that the structure resembles the familiar AdS/CFT and that we can make use of various knowledge of CFTs. However, there is a striking difference from the AdS/CFT
in that there are two asymptotic boundaries in dS, i.e., future and past  infinity. Recently, Cotler and Strominger argued in~\cite{Cotler:2023xku}, motivated by the analogy with the eternal AdS black hole~\cite{Maldacena:2001kr} and the ER=EPR proposal~\cite{Maldacena:2013xja}, that the two CFTs on the boundaries of de Sitter space are entangled, with this entanglement giving rise to the bulk dS geometry---see also~\cite{Narayan:2017xca}.Nevertheless, the nature of this quantum entanglement looks different from that in the eternal AdS black hole because in the dS space the two boundaries are causally influenced. This makes us wonder how this system can be understood in terms of the CFT dual. In this Letter, we address this question by considering the analytic continuation of the boundaries to a Lorentzian torus. After completing the initial draft of this work, we became aware of the independent study~\cite{CS}, which likewise employs the Lorentzian torus to investigate a CFT dual of dS space.\\


\noindent\textbf{2. dS$_3/$CFT$_2$ from Lorentzian torus}. 

\begin{figure}[htbp]
		\centering
		\includegraphics[width=0.48\textwidth]{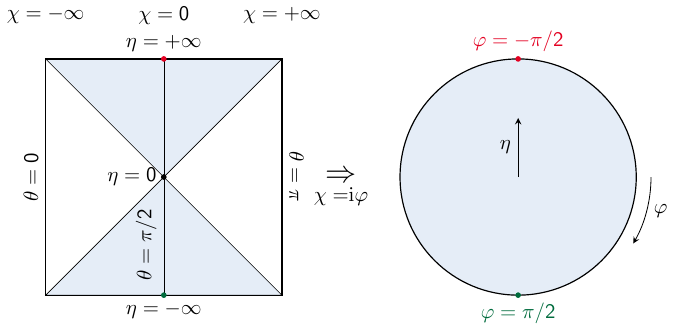}
		\caption{Global structure of dS$_3$ (left) and its Wick rotation into spacetime with two times (right).} 
		\label{fig:setup}
\end{figure}

Three-dimensional dS space (dS$_3$) is described by the hypersurface $-X^2_0+X^2_1+X^2_2+X^2_3=\Ls^2$ with the metric $\diff s^2=-\diff X^2_0+\diff X^2_1+\diff X^2_2+\diff X^2_3$. We can parametrize this in two different ways in terms of the global coordinates $(t,\theta,\phi)$ and in the almost flat slicing
$(\eta,\chi,\phi)$ as 
\begin{align} \label{eq:coords3} 
X_0&= \Ls\sinh\eta\cosh\chi=\sinh t\,,\\
X_1&= \Ls\sinh\eta\sinh\chi=\cosh t\cos\theta\,.\\
X_2&= \Ls \cosh\eta\cos\phi=\cosh t\sin\theta\cos\phi\,,\\
X_3&= \Ls \cosh\eta\sin\phi=\cosh t\sin\theta\sin\phi\,.
\end{align}
The metric of global dS$_3$ is given by
\begin{equation}
\diff s^2 = \Ls^2\left[-\diff t^2+\cosh^2 t \left(\diff\theta^2+\sin^2\theta \,\diff\phi^2\right)\right]\,.
\label{eq:global3}
\end{equation}
On the other hand, the almost flat sliced one looks like
\begin{equation}
\diff s^2 = \Ls^2\left(-\diff\eta^2+\cosh^2\eta\, \diff\phi^2+\sinh^2\eta\, \diff\chi^2\right),
\label{eq:hypsl3}
\end{equation}
where 
$0\leq \theta\leq \pi,  \ 0\leq \phi\leq 2\pi,\ -\infty<\eta,\chi<\infty$.
The coordinate $(\eta,\chi,\phi)$ covers a half of the global dS$_3$ given by $X_0^2-X_1^2>0$, shown as the shaded region in the left panel of Fig.~\ref{fig:setup}.

Although the future and past asymptotic regions,
$t\to+\infty$ and $t\to-\infty$, are disconnected components of the
conformal boundary, they are causally connected through the dS bulk,
and bulk fields can have nonvanishing correlators between them
\cite{Strominger:2001pn,Spradlin:2001nb}. To relate these two
boundaries holographically, we Wick-rotate
\begin{equation}\label{eq:Wickr}
\chi=\iu\left(\vp+\frac{\pi}{2}\right)\,.
\end{equation}
The resulting metric describes a dS$_3$ space with two timelike coordinates (abbreviated as ttdS$_3$),
\begin{equation}\label{twotime}
\diff s^2= \Ls^2\left(
-\diff\eta^2+\cosh^2\eta\,\diff\phi^2
-\sinh^2\eta\,\diff\vp^2\right)\,,
\end{equation}
as shown in the right panel of
Fig.~\ref{fig:setup}---see~\cite{Bars:2000qm,Chen:2025acl,Chen:2025eeh} for other setups with two-time directions. In this geometry, we take the timelike coordinate $\vp$ to be
periodic, $\vp\sim\vp+2\pi$.
The asymptotic boundary at $\eta\to\infty$ is thus a Lorentzian torus, $\mathcal{LT}^2$, parameterized by $(\vp,\phi)\in S^1\times S^1$, by the metric
\begin{equation}\label{eq:dstorus}
\diff s^2_{\mathcal{LT}^2}=-\diff\varphi^2+\diff\phi^2\,,    
\end{equation}
with identifications $\vp\sim\vp+2\pi$, $\phi\sim\phi+2\pi$.

\begin{figure}[htbp]
		\centering
		\includegraphics[width=0.48\textwidth]{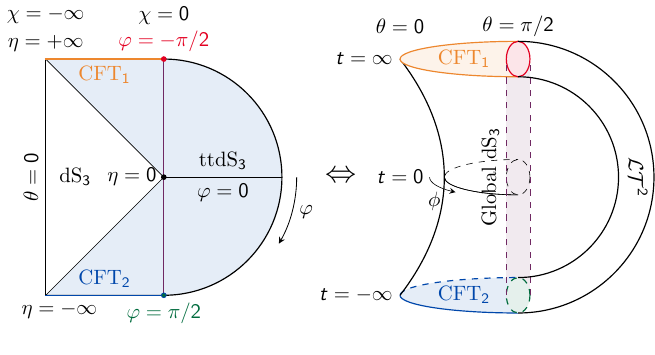}
		\caption{The preparing Hartle--Hawking like state for dS$_3$ via the two-time geometry in the penrose diagram (left) and in the three-dimensional sketch (right).} 
		\label{fig:setupHH}
\end{figure}

Based on these considerations, we propose the duality
\begin{equation}
    \text{gravity on ttdS}_3\Leftrightarrow \text{CFT}_2 \text{ on }\mathcal{LT}^2\,.
\end{equation}
 The dS$_2$ subspace defined by $\theta=\pi/2$ (equivalently,
$\chi=0$) is identified with the $\vp=\pm\pi/2$ slices of the
two-time geometry, as shown in Fig.~\ref{fig:setupHH}. Consequently,
the future and past asymptotic regions of global dS$_3$ are connected
by a cylinder corresponding to one half of $\mathcal{LT}^2$.

To extend this analysis to a dS$_3$ black hole with energy $E$, we have only to replace 
$(\chi,\phi)$ in~\eqref{eq:coords3} and~\eqref{eq:hypsl3} with the rescaled ones $(\lambda\chi,\lambda\phi)$, with $\lambda=\s{1-8\GN E}$ and $\GN$ being the Newton constant. After Wick-rotating~\eqref{eq:Wickr}, we obtain the two-time geometry
\begin{equation}\label{eq:bh2}  
\diff s^2=\Ls^2[-\diff\eta^2+\lambda^2(\cosh^2\eta \,\diff\phi^2-\sinh^2\eta\, \diff\vp^2)]\,.
\end{equation}
The periodicity of $\vp$ and $\phi$ now become $\vp\sim \vp+2\pi/\lambda$ and $\phi\sim \phi+2\pi$. Its cosmological horizon entropy reads
\begin{equation}\label{dSBHent}
 S_\text{dS}=\frac{\pi\lambda}{2\GN}.   
\end{equation}

We now turn to the CFT interpretation. The fact that the future and past boundaries are connected through the two-time cylinder allows us to prepare a quantum state on the two circles at $\chi=0$ by performing the Euclidean path integral of the dual CFT over the cylinder $-\tilde{\beta}_{\dS}/4 \leq \varphi \leq \tilde{\beta}_{\dS}/4$, where $\tilde{\beta}_{\dS}=2\pi/\lambda$. This construction is illustrated in Fig.~\ref{fig:setupHH} for global dS$_3$. The Hilbert space of the dual CFT at $\chi=0$ factorizes as ${\cal H}_1\otimes{\cal H}_2$, where ${\cal H}_1$ and ${\cal H}_2$ correspond to the circles at the future boundary ($t=\infty$) and the past boundary ($t=-\infty$), respectively. We denote the energy eigenstates in these Hilbert spaces by $|n\rangle_1$ and $|n\rangle_2$, with common energy eigenvalue $\Delta_n$.

Notice, however, that this path integral is performed along the temporal direction, since $\varphi$ is a time coordinate. Thus the resulting state turns out to be an unconventional thermofield double (TFD) state,
\begin{equation}\label{TFDt}
\TFD=\sum_{n} e^{-\frac{\beta_{\dS}}{2}\Delta_n}|n\rangle_1|n\rangle_2,
\end{equation}
where the inverse temperature is purely imaginary,
\begin{equation}\label{bds}
\beta_{\dS}=\iu\tilde{\beta}_{\rm dS}=\frac{2\pi\iu}{\lambda}.
\end{equation}
This state can be viewed as a double Wick rotation of the Hartle--Hawking state~\cite{Hartle:1983ai} for the eternal Bañados--Teitelboim--Zanelli (BTZ) black hole~\cite{Banados:1992wn,Banados:1992gq}. Since the two CFTs are related by Lorentzian time evolution on $\mathcal{LT}^2$, they exhibit timelike entanglement in the sense of~\cite{Doi:2022iyj,Doi:2023zaf}. The same form of TFD state as~\eqref{TFDt} was derived in~\cite{Cotler:2023xku} from an analysis of the bulk scalar field and was interpreted as an entangled state in the Hilbert space of two static patches. It would be interesting to work out their direct relations.

To define the conjugate of $\TFD$, we take its Hermitian conjugate while treating $\beta_{\dS}$ as if it were real, 
\be
\DFT=\sum_n e^{-\frac{\beta_{\dS}}{2}\Delta_n}\langle n|_1\langle n|_2. \label{TFDb}
\ee 
Although this is not the actual Hermitian conjugate, this prescription is natural because it reproduces the partition function on $\mathcal{LT}^2$,
$\langle \overline{\text{TFD}}|\text{TFD}\rangle=\sum_{n}e^{-\beta_{\dS}\Delta_n}$, analogous to the prescription employed for the imaginary deformation of the BTZ black hole~\cite{Kawamoto:2025oko,Harper:2025lav}.

To understand the dS entropy, we first note that the central charge dual to the dS$_3$ becomes imaginary in the classical gravity limit
\cite{Maldacena:2002vr,Hikida:2021ese,Hikida:2022ltr}
\begin{equation}
c_{\dS}=\iu\frac{3\Ls}{2\GN}\equiv \iu\ti{c}_{\dS}\,.\label{central}
\end{equation}
If we apply the standard expression of thermal entropy in a two dimensional CFT at finite temperature,
we have 
\begin{equation}\label{dSentca}
S_{\text{th}}=\frac{2\pi^2 c_{\dS}}{3\beta_{\dS}}=\frac{\pi\lambda}{2\GN}\,,
\end{equation}
reproducing the expected dS entropy~\eqref{dSBHent}. 

Since in the dS case we substitute the imaginary values in~\eqref{bds} and~\eqref{central}, let us clarify what is actually being computed. In the CFT dual to dS$_3$, the conformal dimension of a primary operator dual to a bulk scalar field of mass $m$ is expected to be complex~\cite{Strominger:2001pn}
\begin{equation}\label{pmdim}
\Delta^{(\pm)}=1\pm \iu\m\,,\qquad \mu\equiv\sqrt{m^2\Ls^2-1}\,.
\end{equation}
We choose the branch $\Delta^{(+)}$, since we expect that $\Delta^{(+)}$ and $\Delta^{(-)}$ are not independent primaries, in analogy with AdS/CFT. Moreover, we focus on the high-energy regime, where $\Delta^{(+)}\simeq \iu mL$, as these states provide the dominant contribution to the entropy. We therefore parametrize the conformal dimensions as $\Delta_n=\iu\tilde{\Delta}_n$, where $\tilde{\Delta}_n$ is real and positive, with $n$ labeling the primary states. We further assume the density of states in the dS$_3$ dual CFT obeys the Cardy formula,
\begin{equation}\label{cardy}
 \rho(\ti{\Delta})\sim \ex{2\pi\s{\frac{c_{\dS}}{3}\left(\Delta-\frac{c_{\dS}}{12}\right)}}=
\ex{2\pi\s{\frac{\ti{c}_{\dS}}{3}\left(-\ti{\Delta}+\frac{\ti{c}_{\dS}}{12}\right)}},     
\end{equation}
where $\tilde{\Delta}$ is treated as if it were the conformal dimension of a holographic CFT dual to AdS$_3$, up to a power-law prefactor in $\tilde{\Delta}$. Under this assumption, the $\mathcal{LT}^2$ partition function becomes
\begin{equation}\label{zds}
 Z=\int \diff\ti{\Delta}\,\rho(\ti{\Delta}) \ex{-\beta_{\dS}\Delta}
\simeq  \ex{\frac{\pi^2 \ti{c}_{\dS}}{3\ti{\beta}_{\dS}}},  
\end{equation}
where the saddle-point approximation gives $\tilde{\Delta}\simeq-\pi^2 \tilde{c}_{\dS}/(3\tilde{\beta}_{\dS}^2)+\tilde{c}_{\dS}/12=E\Ls$. Evaluating the entropy $\log\rho(\tilde{\Delta})$ at this value reproduces~\eqref{dSentca}, thereby recovering the dS entropy~\eqref{dSBHent}. Here we assumed that the contributions from $\ti\Delta>\ti{c}_{\dS}/12$ can be neglected, as expected from~\eqref{cardy}. 

Another check of our proposal is the agreement between the partition function~\eqref{zds} and the one obtained from the Euclidean gravitational action, $Z_\text{grav}=e^{-I_\text{ren}}$, where $I_\text{ren}$ is the renormalized on-shell action, defined as the limit of the cutoff-regulated action,$I_{\text{ren}}=
\lim_{\eta_\text{c}\to\infty}I$, where the regulated action reads
\begin{equation}\label{eq:onshella}
    I=\frac{1}{16\pi\GN}\left[\int_{\mathcal M}\left[R-\frac{2}{\Ls^2}\right]-2\int_{\partial\mathcal M}\left[K-\frac{1}{\Ls}\right]\right]\,.
\end{equation}
Here, $\int_{\mathcal M}\equiv\int_{\mathcal M}\diff^3 x\sqrt{|g|}$ is the integral on $\mathcal{LT}^2$ and $\int_{\partial\mathcal M}\equiv\int_{\mathcal \partial M}\diff^2 x\sqrt{|\gamma|}$ on its boundary. The last term in~\eqref{eq:onshella} corresponds to the local counterterm. Since $\vp$
has period $2\pi/\lambda$, the $\vp$-cycle closes smoothly at $\eta=0$, and no additional inner boundary term is required.

We now evaluate~\eqref{eq:onshella} in metric~\eqref{eq:bh2}. For the bulk part, we recall that the geometry is locally dS$_3$---which means $R=6/\Ls^2$---and use that $\sqrt{|g|}=\Ls^3\lambda^2\sinh\eta\cosh\eta$. On the other hand, for the boundary part---defined at $\eta=\eta_\text{c}$ by the outward normal $n=\partial_\eta/L$---we have $\sqrt{|\gamma|}=\Ls^2\lambda^2\sinh\eta_\text{c}\cosh\eta_\text{c}$ and $K=
(\cosh^2\eta_\text{c}+\sinh^2\eta_\text{c})/
(\Ls\sinh\eta_\text{c}\cosh\eta_\text{c})$. Then we obtain 
\begin{equation}\label{eq:grav-Z}
\log Z_{\text{grav}}
=-I_{\mathrm{ren}}
=\frac{\pi\Ls\lambda}{4\GN}
=\frac{\pi\ti{c}_{\dS}\lambda}{6}
=\frac{\pi^2c_{\dS}}{3\beta_{\dS}},
\end{equation}
which precisely agrees with the CFT saddle result~\eqref{zds}.\\

\noindent\textbf{3. Correlation functions}. 

The correlation functions on $\mathcal{LT}^2$ with metric~\eqref{eq:dstorus} were pioneered in~\cite{Melton:2023hiq,Melton:2025ecj,Melton:2025jee,Kundu:2025jsm}. We argue that the two point function of a primary operator $\mathcal O_i\equiv \mathcal O(\vp_i,\phi_i)$ of conformal dimension $\Delta$ in dS$_3/$CFT$_2$ is given by
\begin{equation}\label{corCFT}
\langle\mathcal O_1\mathcal O_2\rangle=\frac{1}{\left[\sin\left(\frac{\Delta\varphi+\Delta\phi}{2}\right)\sin\left(\frac{\Delta\varphi-\Delta\phi}{2}\right)-\iu\delta\right]^{\Delta}}\,,
\end{equation}
where $\Delta\varphi=\varphi_2-\varphi_1$, $\Delta\phi=\phi_2-\phi_1$ and $\delta$ is an infinitesimally small positive constant, such that
\begin{equation}\label{signx}
 \frac{1}{(x-\iu\delta)^\Delta}=\frac{1}{|x|^\Delta}\times\begin{cases}
     1 & x>0\,,\\
     \ex{\iu\pi\Delta} &x<0\,.
 \end{cases}   
\end{equation}
Although we follow~\cite{Melton:2023hiq,Melton:2025ecj,Melton:2025jee}, we choose the opposite sign for $\delta$ so that it is compatible with the non-unitary CFT dual to dS$_3$. Notice that in~\eqref{corCFT} and~\eqref{signx}, we have $x>0$ when the two points are timelike separated, whereas $x<0$ when they are spacelike separated.

The bulk two-point function of a scalar field $\Phi_i\equiv\Phi(x_i)$ in the Hartle--Hawking vacuum is given by~\cite{Bousso:2001mw} 
\begin{equation}\label{eq:phiphi}
\langle \Phi_1\Phi_2\rangle
=
\frac{\sinh[\mu(\pi-D_{12})]}{\sinh(\pi\mu)\,\sin D_{12}}.
\end{equation}
Here $D_{12}$ is the geodesic distance between the spacetime points $x_1$ and $x_2$ in dS$_3$\begin{equation}
\cos D_{12}=
\omega_{12}\,\cosh t_1\cosh t_2-\sinh t_1\sinh t_2\,,
\label{geodesicL}
\end{equation}
where $\omega_{12}\equiv\mathbf{\Omega}_1\cdot\mathbf{\Omega}_2$, with $\mathbf{\Omega}$ denoting the embedding vector of a point on the unit $S^2$, satisfying $\mathbf{\Omega}\cdot\mathbf{\Omega}=1$.

If we consider two future boundary points $t_1=t_2=t_\infty\to \infty$, then $D_{12}\simeq \pi+\iu\log\left[\ex{2t_\infty}(1-\omega_{12})/2\right]$, and, from~\eqref{eq:phiphi}, we obtain
\begin{align}
 \la\Phi_1\Phi_2\lb_{++}\simeq &\frac{\sin\left[\mu\log\left[\frac{\ex{2t_\infty}}{2}(1-\omega_{12})\right]\right]}{\sinh(\pi\mu)\frac{\ex{2t_\infty}}{4}(1-\omega_{12})}\no
=&\frac{\iu}{\sinh(\pi\mu)}\left[\frac{\ex{2t_\infty}}{2}(1-\omega_{12})\right]^{-\Delta^{(+)}}\no
&-\frac{\iu}{\sinh(\pi\mu)}\left[\frac{\ex{2t_\infty}}{2}(1-\omega_{12})\right]^{-\Delta^{(-)}}\,.  \label{ppcor}
\end{align}
Note that the past infinity one $\la \Phi_1\Phi_2\lb_{--}$ is identical.

On the other hand, if we take $x_1$ to be the future infinity, while $x_2$ to be the past infinity, $t_1=-t_2=t_\infty\to \infty$, we find
$D_{12}\simeq \iu\log\left[\ex{2t_\infty}(1+\omega_{12})/2\right]$, which leads to
\begin{align}
\la\Phi_1\Phi_2\lb_{+-}\simeq&\frac{
\ex{\pi \mu}\ex{-\mu D_{12}}-\ex{-\pi \mu}\ex{\mu D_{12}}}{\iu\sinh(\pi \mu) \frac{\ex{2t_\infty}}{2}(1+\omega_{12})}\no
=&-\frac{\iu \ex{\pi\mu}}{\sinh(\pi\mu)}\left[\frac{\ex{2t_\infty}}{2}(1+\omega_{12})\right]^{-\Delta^{(+)}}\no
&
+\frac{\iu \ex{-\pi\mu}}{\sinh(\pi\mu)}\left[\frac{e^{2t_\infty}}{2}(1+\omega_{12})\right]^{-\Delta^{(-)}}\,.  \label{mmcor}
\end{align}

From the CFT perspective, in the asymptotic limits $t\to \infty$ and $t\to -\infty$, we write
\begin{equation}
\Phi \simeq
\sum_{\sigma=\pm}
\frac{\iu^{\sigma\frac{1}{2}}
\ex{-\left(t_\infty+\frac{\iu\pi}{2}\right)\Delta^{(\sigma)}}}{\sqrt{\sinh\pi \mu}}\times
\begin{cases}
\mathcal O_\sigma(\mathbf\Omega)& t\rightarrow\infty\,,\\
\hat{\mathcal O}_\sigma(\mathbf\Omega)& t\rightarrow-\infty\,,
\end{cases}
\end{equation}
where $\mathcal O_{\pm}(\mathbf \Omega)$ and $\hat{\mathcal O}_{\pm}$ denote the CFT operators on $\mathcal{LT}^2$ inserted at $\varphi=-\pi/2$ and $\varphi=\pi/2$, respectively, with conformal dimensions $\Delta^{(\pm)}$ given in~\eqref{pmdim}.
Substituting this expansion into the asymptotic bulk two-point function~\eqref{ppcor} then gives
\begin{equation}
\la \mathcal O_{\pm}(\mathbf\Omega_1)\mathcal O_{\pm}(\mathbf\Omega_2)\lb=\frac{\ex{\pi \iu\Delta^{(\pm)}}}{\left(\frac{1-\omega_{12}}{2}\right)^{\Delta^{(\pm)}}},
\end{equation}
which reproduces the CFT result~\eqref{corCFT} at $\vp_1=\vp_2=-\pi/2$. On the other hand,~\eqref{mmcor} leads to
\begin{equation}
\la \mathcal O_{\pm}(\mathbf\Omega_1)\hat{\mathcal O}_{\pm}(\mathbf\Omega_2)\lb=\frac{1}{\left(\frac{1+\omega_{12}}{2}\right)^{\Delta^{(\pm)}}},
\end{equation}
agreeing with the CFT result~\eqref{corCFT} at $\vp_1=-\pi/2$ and $\vp_2=\pi/2$.
The divergence of the correlator at $1+\omega_{12}=0$ signals strong entanglement between a point $\vp$ on the first CFT and its antipodal point $\vp+\pi$ on the second CFT, within our timelike TFD state. Such a role of the antipodal map---see also~\cite{Witten:2001kn,Harlow:2023hjb,Doi:2024nty,Goodhew:2024eup} for other aspects---is a peculiar feature that is absent in the standard TFD state. Extending the above analysis to the dS$_3$ black hole is straightforward: one simply replaces $(\vp,\phi)$ with the rescaled coordinates $(\lambda\vp,\lambda\phi)$. 

This setup allows us to perform an explicit computation of the pseudoentropy associated with the reduced transition matrix $\tau_A$, defined as $S_A=-\text{tr}(\tau_A\log\tau_A)$~\cite{Nakata:2021ubr}. Here $\tau_A$ is defined from the wave-function of the total system by tracing out the complement of $A$ and is non-hermitian in the dS/CFT as the dual CFT is non-unitary~\cite{Doi:2022iyj,Doi:2023zaf}---see also, e.g.,~\cite{Mollabashi:2020yie,Mollabashi:2021xsd,Hikida:2022ltr,Narayan:2022afv,Narayan:2023ebn,Narayan:2023zen,Anastasiou:2025rvz,Anastasiou:2026bbf,Narayan:2026wzp}. Choosing the subsystem $A$ to be the interval $[\phi_1,\phi_2]$ at $\vp=-\ti{\beta}_{\dS}/4$---or equivalently $\theta=\pi/2$---we can compute $S_A$ via the replica method~\cite{Calabrese:2004eu,Calabrese:2009qy} by evaluating the two-point function of twist operators $\sigma_n$ and anti-twist operators $\bar\sigma_n$, restricted to the slice $\vp=-\ti{\beta}_{\dS}/4$, that is,
\begin{equation}
S_A=\lim_{n\to 1}\frac{1}{1-n}\log\la \sigma_n(\phi_1)\,\bar{\sigma}_n(\phi_2)\lb\Big|_{\vp=-\ti{\beta}_{\dS}/4}\,.
\end{equation}
By plugging the conformal dimension $\Delta=\iu\ti{c}_{dS}(n-1/n)/12$ into~\eqref{corCFT}, we obtain
\begin{equation}
S_A=\frac{\iu \ti{c}_{\dS}}{3}\log \frac{2\sin\left(\lambda\Delta\phi/2\right)}{\lambda\ep}+\frac{\pi \ti{c}_{\dS}}{6}\ ,
\end{equation}
where we have compensated for the cutoff $\ep$ dependence to restore the canonical form. The holographic pseudoentropy is given by $S_A=D_{12}/(4\GN)$, analogous to the holographic entanglement entropy~\cite{Ryu:2006bv,Ryu:2006ef,Hubeny:2007xt}, where $D_{12}$ is the geodesic length in dS$_3$, given by~\eqref{geodesicL}, connecting the two endpoints of $A$. The CFT on $\mathcal{LT}^2$ thus reproduces the gravity result for $S_A$.\\

\noindent\textbf{4. Emergence of dS from path-integral optimization}. 

Our $\mathcal{LT}^2$ formulation allows us to understand how dS$_3$ space emerges from the dual CFT, extending the path-integral optimization for AdS$_3$/CFT$_2$~\cite{Caputa:2017urj,Caputa:2017yrh,Boruch:2020wax,Boruch:2021hqs}---a continuum version of emergent-space-from-tensor-network ideas~\cite{Swingle:2009bg,Nozaki:2012zj,Pastawski:2015qua,Czech:2015kbp,Hayden:2016cfa,Milsted:2018yur,Takayanagi:2018pml}---to dS$_3$/CFT$_2$, as shown below.

Briefly, in AdS$_3$/CFT$_2$ the CFT vacuum is computed by a Euclidean path-integral with metric $\diff s^2=\ex{2w}(\diff \tau^2+\diff x^2)$, where $w=w(\tau,x)$ is the Weyl factor, normalized so each lattice site occupies unit area. The flat-space path-integral corresponds to $e^{2w}=1/\ep^2$ with the lattice constant $\ep\to0$. Weyl invariance of the CFT implies that changing $w$ only rescales the wave functional by $\ex{S_\text{L}}$, with Liouville action
\begin{equation}
S_{\text L}=\frac{c}{24\pi}\int^{-\ep}_{-\infty} \diff\tau \int \diff x \left[(\de_\tau w)^2+(\de_x w)^2+\ex{2w}\right].
\end{equation}
Minimizing $S_{\text L}$---interpreted as the computational complexity of the lattice---subject to $\ex{2w}=1/\ep^2$ at $\tau=-\ep$, gives the Liouville equation $(\de_\tau^2+\de_x^2)w=e^{2w}$, solved by $\ex{2w}=\tau^{-2}$. This explains the emergence of hyperbolic-plane slice $\diff s^2=(\diff \tau^2+\diff x^2)/\tau^2$ of Poincar\'e AdS$_3$.

Now we extend this setup to dS$_3/$CFT$_2$. We consider the Weyl transformation of the Lorentzian cylinder whose path-integral prepares our timelike TFD state~\eqref{TFDt}, $\diff s^2=\ex{2w}\diff s^2_{\mathcal{LT}^2}$, with periodicities $-\ti{\beta}_{\dS}/4\leq \vp\leq\ti{\beta}_{\dS}/4$ and $0\leq \phi\leq 2\pi$, respectively.

The wave functional is proportional to $\propto \ex{\iu S_{\text L}}$, where
\begin{equation}
S_{\text L}=\frac{\iu\ti{c}_{dS}}{24\pi}\int \diff \vp \,\diff\phi\left[-(\de_\vp w)^2+(\de_\phi w)^2-e^{2 w}\right]\,.
\end{equation}
Since $\iu S_{\text L}$ is real-valued and thus can be minimized assuming the translational invariance 
$\de_\phi w=0$ as the central charge $\iu \ti{c}_{\text{dS}}$ is imaginary. This leads to the solution 
\begin{equation}
\ex{2w}=\frac{4\pi^2}{\beta^2\cos^2\left(\frac{2\pi \vp}{\ti{\beta}_{\dS}}\right)}.
\end{equation}
This coincides with the metric of the dS$_2$ slice $\chi=0$  of the dS$_3$ or its black hole deformation---see Fig.~\ref{fig:setupHH}.
Thus, the dS$_2$ emerges by optimizing the Lorentzian cylinder.\\

\noindent\textbf{5. Higher-dimensional generalization}.

We can extend the duality presented in dS$_3$/CFT$_2$ to higher dimensions. To that end, consider the $(d+2)$-dimensional metric $\diff s^2=-\diff X_0^2+\sum_{i=1}^{d+1}\diff X_i^2$, in which dS$_{d+1}$ is given by the embedding $-X_0^2+\sum_{i=1}^{d+1}X_i^2=\Ls^2$. As before, we parametrize this using global coordinates $(t,\theta_1,\ldots,\theta_d)$, related to the embedding coordinates as
\begin{align}
    \frac{X_0}{\Ls}=&\sinh t\,,\\
    \frac{X_i}{\Ls}=&\cosh t\times\begin{cases}
    \cos\theta_i\prod_{j=1}^{i-1}\sin\theta_j \quad & i=1,\ldots, d\,,\\
    \prod_{j=1}^{d}\sin\theta_j & i=d+1\,.
    \end{cases}
\end{align}
The ranges of these coordinates are such that $-\infty<t<+\infty$, $0\leq\theta_j\leq\pi$ ($j=1,\ldots,d-1$) and $0\leq\theta_d<2\pi$. Notice that $\theta_d$ plays the role of the azimuthal angle; in the notation of~\eqref{eq:coords3}, one identifies $\theta_d \equiv \phi$. In this writing, the higher-dimensional version of the global coordinates presented in~\eqref{eq:global3} reads
\begin{equation}
    \diff s^2=\Ls^2\left(-\diff t^2+\cosh^2t\,\diff\Omega_{d}^2\right)\,,
\end{equation}
where $\diff\Omega_d^2 =
\sum_{i=1}^{d} \left(\prod_{j=1}^{i-1}\sin^2\theta_j\right)\diff\theta_i^2$
is the metric of the $d$-sphere.
We now extend the coordinate transformation introduced in~\eqref{eq:coords3} to the higher-dimensional coordinates $(\eta,\chi,\theta_1,\ldots,\theta_{d-1})$ according to
\begin{align}
\frac{X_0}{\Ls} &= \sinh\eta\cosh\chi\,,\qquad \frac{X_1}{\Ls} =\sinh\eta\sinh\chi\,,\\
\frac{X_i}{\Ls} &=\cosh\eta\times
\begin{cases}
\cos\theta_{i-1}\prod_{j=1}^{i-2}\sin\theta_j
& i=2,\ldots,d\,,\\
\prod_{j=1}^{d-1}\sin\theta_j
& i=d+1\,,
\end{cases}
\end{align}
with the ranges given by $0\leq\eta<\infty$, $-\infty<\chi<\infty$, together with $0\leq\theta_j\leq\pi\quad (j=1,\ldots,d-2)$, 
$0\leq\theta_{d-1}<2\pi$. Substituting these expressions into the embedding metric, one obtains
\begin{equation}
\label{eq:dSd-chi}
\diff s^2=
\Ls^2\left(
-\diff\eta^2
+\sinh^2\eta\, \diff\chi^2
+\cosh^2\eta\, \diff\Omega_{d-1}^2
\right)\,.
\end{equation}
Now, performing the Wick rotation~\eqref{eq:Wickr} we obtain
\begin{equation}\label{eq:dSd}
\diff s^2
=
\Ls^2\left(
-\diff\eta^2
-\sinh^2\eta\, \diff\varphi^2
+\cosh^2\eta\, \diff\Omega_{d-1}^2
\right) ,
\end{equation}
The asymptotic boundary at $\eta\to\infty$ is
$S^1_{\varphi}\times S^{d-1}$, a Lorentzian manifold with the $S^1_{\varphi}$
circle being timelike. This generalizes the $\mathcal{LT}^2$ in the $d=2$ case.

Exactly as before, restricting the path integral to
$-\pi/2\leq \varphi\leq\pi/2$ with
$\Omega_{d-1}\in S^{d-1}$ defines two copies of the CFT$_d$ at
$\varphi=\mp\pi/2$, and the bulk path integral prepares the TFD state (\ref{TFDt}) and (\ref{TFDb}),
where $\Delta_n$ now labels the energy eigenvalues of the CFT$_d$ quantized on
$S^{d-1}$ and we have again $\beta_{\dS}=2\pi \i$.

Again, the horizon is located at the $\eta=0$ slice. Thus, the Gibbons--Hawking entropy
\begin{equation}
S_{\dS}
=\frac{\Ls^{d-1}}{4\GNd}\int\diff\Omega_{d-1}=\frac{\Ls^{d-1}\Omega_{d-1}}{4\GNd}\,,  \label{hdime}
\end{equation}
where $\Omega_{d-1}=2\pi^{d/2}/\Gamma(d/2)$ is the volume of the $(d-1)$-dimensional unit sphere $S^{d-1}$. 

To see the CFT dual, consider the AdS-Schwarzschild black hole in $d+1$ dimensions,
$\diff s^2=-f(r)\diff\tau^2+f(r)^{-1}\diff r^2+r^2\diff\Omega^2_{d-1}$,
where $f(r)=1+r^2/\Ls_\text{AdS}^2-\mathsf M/r^{d-2}$ and $\mathsf M$ is an integration constant related to the ADM mass $M$ through $\mathsf M=16\pi\GN^{(d+1)}M/[(d-1)\Omega_{d-1}]$~\cite{Arnowitt:1960es,Arnowitt:1960zzc,Arnowitt:1961zz}. Via the analytic continuation $\Ls_{\rm AdS}=\iu\Ls$, the thermodynamics of the AdS black hole~\cite{Witten:1998zw} yields the dS counterparts of the density of states $\rho$ and the energy $\Delta$, i.e.,
\begin{align}
& \log \rho\simeq S_{\dS}\cdot x^{d-1}\,,\\
&\beta_{\dS}\Delta\simeq \frac{(d-1)\ti{\beta}_{\dS}}{4\pi}(x^{d}-x^{d-2})\cdot S_{\dS}\,,
\end{align}
where $x=r_+/\Ls$ is the black hole in units of the dS radius. The partition function~\eqref{zds} can then be written as
$Z\simeq \int \diff x\, \rho\, e^{-\beta_{\dS}\Delta}$. Its saddle point at $\beta_{\dS}=2\pi\iu$ occurs at $x=1$, yielding the entropy~\eqref{hdime} in the form $\log\rho\simeq S_{\dS}$.\\

\noindent\textbf{6. Discussions.}

In this article we proposed a new formulation of the dS/CFT correspondence where the dual CFT lives on $\mathcal{LT}^2$. 
We mainly focused on the three-dimensional case and provided several evidence from various viewpoints. Clearly, there are many more aspects to be explored further, which include lower and higher dimensional cases as well as many other observables in the dS/CFT. 
We intend to address these issues soon~\cite{dSlong}.\\

\noindent\textbf{Acknowledgments}:
We are very grateful to Jordan Cotler and Andrew Strominger for valuable comments on the preliminary draft of this article. We also thank Kanato Goto, Xiao-liang Qi and Yu-ki Suzuki for useful discussions. This work is supported by MEXT KAKENHI Grant-in-Aid for Transformative Research Areas (A) through the ``Extreme Universe'' collaboration: Grant No.~21H05187. K.F. is also supported by Grant-in-Aid for JSPS Fellows No.~26KJ1554. M.K. is  supported by Grant-in-Aid for JSPS Fellows No.~26KJ1545. J.M. is supported by the Beatriu de Pinós fellowship BP 2024 00033 of the Agència de Gestió d'Ajuts Universitaris i de Recerca, Generalitat de Catalunya. K.S. is supported by Grant-in-Aid for JSPS Fellows No.~25KJ1498.  T.T. is also supported by JSPS Grant-in-Aid for Scientific Research (B) No.~25K01000.


\bibliography{dSTEE.bib}


\end{document}